# Title: Multi-Modal Neuroimaging Analysis and Visualization Tool (MMVT)

**Authors:** O. Felsenstein[c&], N. Peled[a,b&], E. Hahn[b], A. P. Rockhill[b,i], D. Frank[h], A M. Libster[g], Y. Nossenson[k], L. Folsom[d], T. Gholipour[f], K. Macadams[a], N. Rozengard[a], A. C. Paulk[b,d], D. D. Dougherty[b,e], S. S. Cash[b,d], A. S. Widge[e,j], M. Hämäläinen[a,b], S. Stufflebeam[a,b]

(a) Athinoula A. Martinos Center. for Biomedical Imaging, Department of Radiology, Massachusetts General Hospital, Charlestown, MA;

(b) Harvard Medical School, Boston, MA

(c) Gonda Multidisciplinary Brain Research Center, Bar-Ilan University, Ramat-Gan, Israel;

(d) Neurology, Massachusetts General Hospital, Boston, MA;

(e) Psychiatry, Massachusetts General Hospital, Boston, MA;

(f) The George Washington University, Washington, DC

(g) University California San Diego, department of medicine, San Diego, California, USA

(h) Universidad Politécnica de Madrid

(i) Present Address: University of Oregon, Eugene, OR

(j) Present Address: Department of Psychiatry, University of Minnesota, Minneapolis

(k) University of Massachusetts Lowell, Lowell MA

[&] Both authors contributed equally to this manuscript.

**Corresponding Author:** N. Peled, Athinoula A. Martinos Center. for Biomedical Imaging, Department of Radiology, Massachusetts General Hospital, Charlestown, MA 02129

Tel.: +1 (617) 959-0457

E-mail address: npeled@mgh.harvard.edu

Multi-Modal Neuroimaging Analysis and Visualization Tool (MMVT)

**Abstract**

Sophisticated visualization tools are essential for the presentation and exploration of human neuroimaging data. While two-dimensional orthogonal views of neuroimaging data are conventionally used to display activity and statistical analysis, three-dimensional (3D) representation is useful for showing the spatial distribution of a functional network, as well as its temporal evolution. For these purposes, there is currently no open-source, 3D neuroimaging tool that can simultaneously visualize desired combinations of MRI, CT, EEG, MEG, fMRI, PET, and intracranial EEG (i.e., ECoG, depth electrodes, and DBS). Here we present the Multi-Modal Visualization Tool (MMVT), which is designed for researchers to interact with their neuroimaging functional and anatomical data through simultaneous visualization of these existing imaging modalities. MMVT contains two separate modules: The first includes complete stand-alone pre-processing pipelines, from raw data to statistical maps. The second module is an add-on to the open-source, 3D-rendering program Blender. It is an interactive graphical interface that enables users to simultaneously visualize multi-modality functional and statistical data on the cortex and in subcortical surfaces as well as the activity of invasive electrodes. This tool also enables highly accurate 3D visualization of neuroanatomy, including the location of invasive electrodes relative to brain structures. Each of the modules and module features can be integrated, separate from the tool, into existing data pipelines. MMVT leverages open-source software to build a comprehensive tool for data visualization and exploration This gives the tool a distinct advantage in both clinical and research domains as each has highly specialized visual and processing needs.

## Introduction

Researchers and clinicians use multimodal imaging in basic and clinical research (Nowell et al., 2015). Using modalities with complementary physiological sensitivities provides cross-validation and facilitates characterization of interactions and their causality (Biessmann et al., 2011). While multimodal approaches offer the possibility of minimizing the limitations of each modality (in terms of spatial and temporal resolution), there are several challenges that arise from data collection and data analysis (Lahat et al., 2015). Initially, gathering sufficient data from several unimodal methods was a major obstacle. This problem has been addressed by the collection of large-scale databases that store and share multiple data modalities for groups of subjects, such as the Human Connectome Project (Daducci et al., 2012; Hodge et al., 2016) and the multi-site clinical trial Establishing Moderators and Biosignatures of Antidepressant Response in Clinical Care (EMBARC) project (Trivedi et al., 2016).

The preprocessing stage of multimodal research includes challenges such as removing artifacts introduced during simultaneous recording or imaging processes. For example, the MRI signal can get affected by susceptibility artifacts produced by EEG (Uludağ and Roebroeck, 2014) and EEG recorded during MRI acquisition will get artifacts from both the switching gradient fields and the ballistocardiogram (BCG). This challenge can be effectively addressed using modality-specific open-source software (Fischl, 2012; Gramfort et al., 2014, 2013; Savio et al., 2017). Additionally, the coordinate systems of the different modalities must be fused to a joint space. This challenge originates from the different spatial resolutions, as well as differences in origin and reference points, as defined and measured by the different instruments. A common solution for this challenge is to transform the coordinate system of one modality to the coordinate system of the other. Similarly, multiple co-registration tools exist, however most such tools are limited to registration between two specific modalities such as MRI and MEG.

The integration of multimodal data analysis can be conducted at multiple levels, as been comprehensively reviewed (Calhoun and Sui, 2016). While multimodal datasets can be analyzed separately and integrated subsequently as in the presurgical assessment of patients with epilepsy, in order to extract as much information as possible, it is, at least theoretically, best to analyze the multimodal dataset jointly. Generally, the development of new methods requires constant visualization both to gain initial insights and for quality control. As the complexity of analysis increases, visualizing the data in a clear and interpretable way becomes more complex, especially as several modalities are analyzed and inspected simultaneously.

Another challenge is to visualize data naturally and intuitively. While neural data is recorded in 3D space, most data visualization is done using 2D slices. Such visualization may originate from the challenge of presenting volumetric and subcortical data efficiently. However, 2D visualizations prevent a full understanding of complex formations, tracts and structures, and the exact origin of signals in space (often not located on a common 2D plane). For example, it has been shown in combined fMRI and brain stimulation studies that stimulation affects a widespread network, involving interconnected brain regions (Basu et al., 2019; Hartwigsen et al., 2013; Turi et al., 2012; Volman et al., 2011; Widge et al., 2019).

One significant challenge in visualization is controlling the amount of information presented. For example, a good visualization needs to provide the user with ways to overcome occlusions. Occlusions can arise from using a specific point of view, visualizing subcortical and cortical structures simultaneously, or by using a surface type that is not suitable (for example presenting activity in sulci using a pial surface). Additional challenges arise from an excessive amount of information; one should aim to reduce complexity whenever possible. Solutions include enabling exploring electrodes one by one in a structured way or masking specific regions of interest according to a chosen parameter.

Finally, multimodal research faces challenges creating connections between communities employing each modality and their paradigms and practices. Each community uses its own software, file formats, atlases, and common parcellations, template brains, coordinate systems and so on. For example, comparing new findings with the prevalent literature is burdened by the need for multiple conversions and adjustments. While there are endeavors to solve each challenge separately, each requires its own software and know-how which takes time and effort (Chau and McIntosh, 2005; Esteban et al., 2019;

Gorgolewski et al., 2016; Savio et al., 2017). A unified solution is needed for bridging the gaps and improving communication between communities.

The central purpose of our new Multi-Modal Visualization Tool (MMVT) is to act as a unified platform for exploring neuroimaging data. Functionally, MMVT links existing specialized analytical tools; it does not replace them. It offers a novel approach to visualization and data exploration, but capitalizes on existing unimodal tools and provides a single platform to access them. Accordingly, it can be used alongside existing analysis pipelines or integrated into the experimental design. Unlike self-contained analysis packages with visualization capabilities, MMVT does not focus on the analysis of data. Instead, it focuses on creating a means of simultaneous exploration of data resulting from multiple neuroimaging modalities. This enables researchers and clinicians to dynamically explore multimodal neuroimaging, both on the individual and group levels. Clinicians are provided with means of visually representing separate, clinically relevant, tools for diagnostics and prognostics. Researchers gain flexibility that allows for intuitive visualization of complex relationships in space, time, and frequency.

Below, we outline the general framework of MMVT and how it interfaces with other software. These sections include the process of importing data from different neuroimaging modalities to MMVT. Following this, we present case studies that illustrate different facets of multimodal neuroimaging in MMVT. We conclude by highlighting how the research of brain stimulation together with neuroimaging and electrophysiological modalities could benefit from using MMVT.

## Methods

In this section, we describe the MMVT framework and software components (Section 2.1). We then introduce the pipeline; specifically, creating and importing a new subject to MMVT and describing how data can be imported from standard analysis software into MMVT (Section 2.2). In section 2.3, we present the different ways to interact with the interface (e.g. utilizing Jupyter Notebook, Python code) and exporting options (e.g. figures, movies, pdf reports, files for sharing results). Finally, we demonstrate the utility of the software given diverse datasets included in the Result section (Section 2.4).

### *2.1 MMVT implementation*

MMVT is open-source software, implemented in Python, that can be deployed on the three major operating systems - Windows, Mac, and Linux. MMVT is composed of two main parts: (1) Graphical modules that include both the visualization components and the graphical user interface (GUI) and (2) pre-processing modules for anatomical and functional data.

The graphical modules of MMVT include both the visualization constituents and the graphical user interface (GUI). These modules are implemented as add-ons in Blender, which is a free and open-source 3D cross-platform creation suite (Roosendaal and Wartmann, 2000). We capitalize on Blender visualization capabilities by creating a set of shaders that enable users to render highly detailed transparent, semi-transparent, and opaque brains while removing visual clutter based on the neural activity presented.

The pre-processing modules encompass a wide range of steps from signal analysis through data transformation and across imaging modalities, formats, and data types. The pre-processing modules include comprehensive pipelines from the creation of the anatomical surfaces through reading the raw or processed output files (of various acquisition modalities), cleaning data, to higher-level analysis such as functional connectivity. Pre-processing pipelines are implemented as stand-alone modules and are built on public libraries and software, such as MNE-Python (Gramfort et al., 2014, 2013),

Freesurfer (Dale et al., 1999; Fischl, 2012; Fischl et al., 2004, 1999), Nibabel (Brett et al., 2016) and others. The pre-processing functions can be run from the command line, Python graphical interface (PyQT5) or directly from the main graphical module. This feature is particularly notable as it allows for partial use or "hacked" coding of the MMVT software; thus, allowing users to integrate specific features - such as multi-modal co-registration - into existing pipelines. The output files can be imported into MMVT or used as stand-alone by the user. *Figure 1* presents the graphical wizard for the preprocessing stages of reconstructing the anatomical structures.

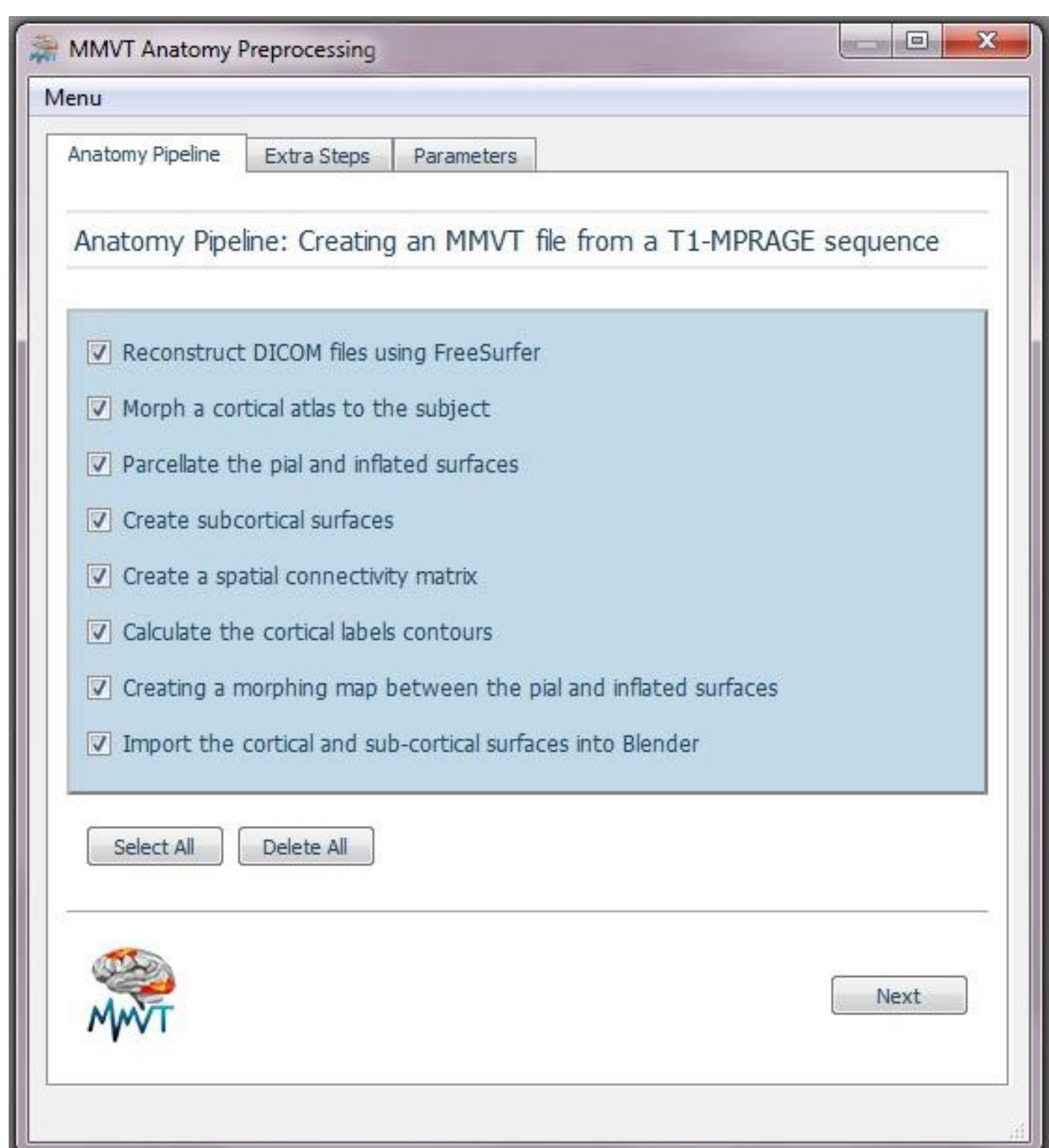

**Figure *1.* The anatomy pipeline GUI, built as a wizard**

*2.2 Subject creation pipeline*

If the subject has a CT scan, it can be co-registered with the MRI. Furthermore, the user can preprocess and import functional data (EEG, MEG, fMRI) and invasive electrodes. *Figure 2* presents the pipeline processing steps. The first step is the creation of the anatomical model to be used for the analysis. Users can decide between various template brain surfaces or individual subject brain models. If individual anatomy is employed, the first step is to run the FreeSurfer MRI reconstruction pipeline. The anatomy processing pipeline converts the cortical surfaces to a Blender-compatible format, creates the subcortical surfaces, sub-parcellates the cortex to different labels according to a selected atlas, and creates the morphing function between the pial and inflated surfaces. Next, the data from the functional modalities of interest can be processed using the MMVT preprocessing pipelines or with external tools. MMVT fully supports importing data from MNE (Gramfort et al., 2014, 2013), fsFast (FreeSurfer Functional Analysis Stream), and SPM (Penny et al., 2011). There is some support for importing data from FieldTrip (Oostenveld et al., 2011), 3D Slicer (Pieper et al., 2004), Brainstorm (Tadel et al., 2011), and BIDS (Niso et al., 2018). Users can easily import different types of data into MMVT, by converting the data to one of the following file types: DICOM, NIfTI, "fif", "img", or Python numpy array. For an in-depth description of the preprocessing stages, see the sections below and *Figure 2*. The last step is to load the subject's data and brain surfaces to the subject file.

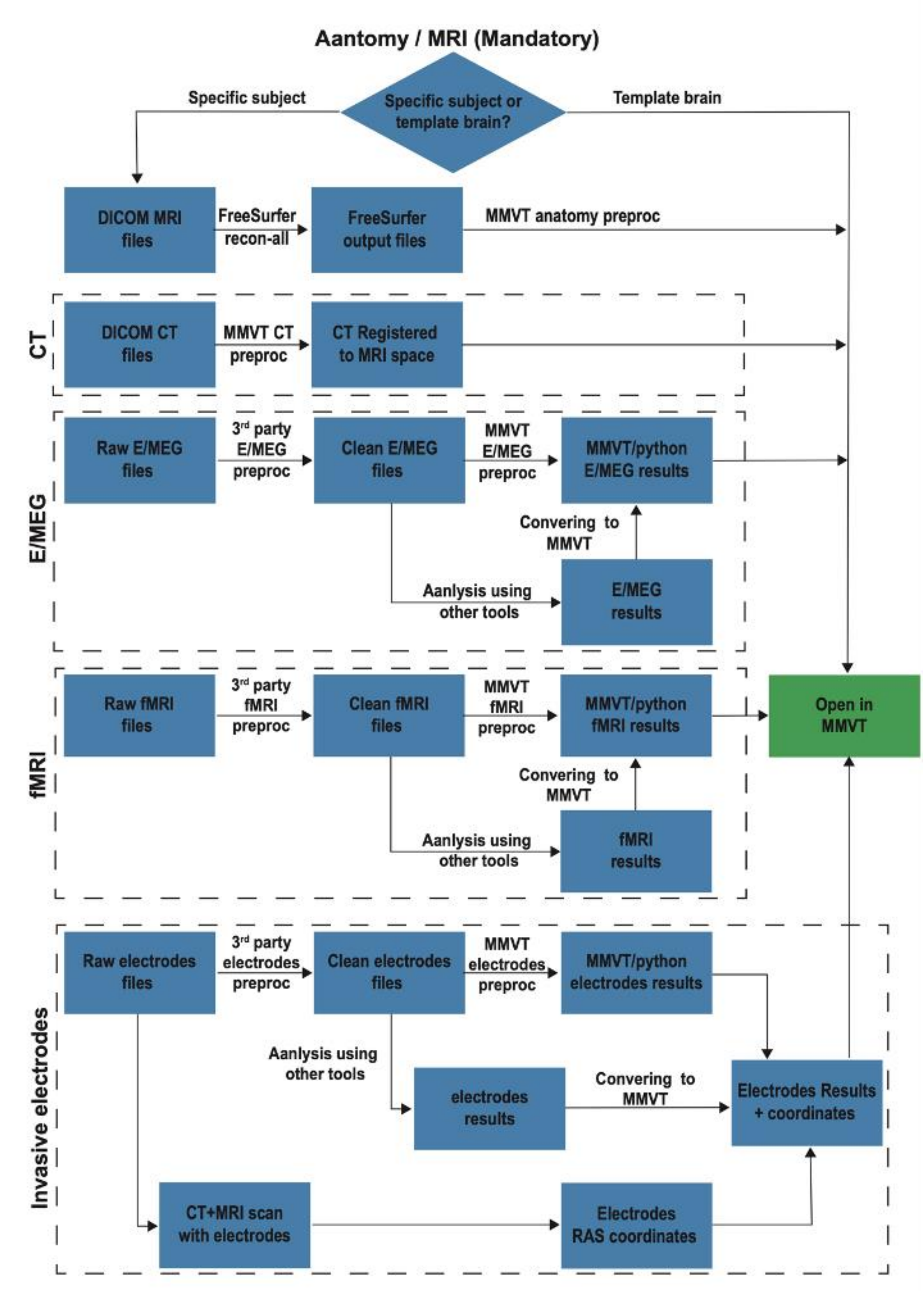

**Figure *2*. MMVT pre-processing flow-chart**

**MRI and CT:** The user has the option to use a template brain, e.g., to visualize group analysis results, see *Figure 2*, parts (1) and (2). The MMVT template brain (including fsaverage, the Freesurfer average brain, or MNI's colin27 (Holmes et al., 1998)) can be downloaded from our website (mmvt.mgh.harvard.edu). For subject-specific brains, the user can import the subject's MRI and reconstruct it using MMVT. The MRI pipeline is based on existing free software, FreeSurfer, developed for the MR imaging preprocessing and analysis (Dale et al., 1999; Fischl, 2012; Fischl et al., 2004, 2002, 1999; Postelnicu et al., 2009). Preprocessing of the structural MRI data includes the anatomical reconstruction (generated from the T1-MPRAGE) by using the FreeSurfer recon-all module. Aside from the built-in steps in the FreeSurfer, the necessary steps are as follows: parcellation of the pial and inflated surfaces to cortical labels, segmentation of the subcortical surfaces via the marching cubes algorithm (Lorensen and Cline, 1987), creation of a morphing function between the pial and the inflated surfaces via Morph target animation (Liu, 2009), determining the neighboring vertices on the cortical surfaces, and the importation of the objects into Blender. This last step will create the final Blender file that will include the cortical and subcortical structures as well as the inflated surfaces.

In addition to MRI, a subject's CT scan can be also be incorporated. A relevant application of this is where a CT scan is needed for surgical planning. An example of a use case is localization of implanted electrodes in epilepsy patients. The CT data can be registered to the subject's MRI space using the MMVT CT pre-processing pipeline, which is primarily a wrapper around the FreeSurfer registration method. This method employs robust statistics to detect and remove outliers from the registration and therefore leads to highly accurate registrations Moreover, a merged NIfTI file is created by combining elements from both the MRI and the CT that can be plotted in the MMVT slices viewer.

**EEG and MEG:** The EEG and MEG pre-processing module mostly employs MNE-Python (Gramfort et al., 2013) function calls invocations of command-line utilities, *Figure 2*, part (3). The pipeline starts with functions that clean and read raw data and concludes with steps of calculating average evoked responses over cortical labels and connectivity of cortical labels. In addition to wrapping existing modules in MNE-Python, further steps were implemented to pre-process the EEG/MEG data before importing it into Blender. For example, MMVT can create a 3D mesh for the EEG cap and MEG "helmet" for plotting the potentials over time and finding clusters of functional regions of interest (ROIs) in a source estimate by using the connectivity by using the connectivity matrix of the cortical surfaces (Gramfort et al., 2013; Pearce, 2005).

**fMRI and PET:** The fMRI/PET pipeline is designed to clean raw fMRI and/or PET data; starting from DICOM files, up to calculating a contrast, group average, and connectivity, *Figure 2*, part (4). Additional supported features include: morphing volumetric files to and from a template brain and between subjects, projection of a volumetric file on the cortical and subcortical surfaces, averaging of a contrast file over cortical labels and subcortical regions, finding clusters of activation, (Gramfort et al., 2013; Pearce, 2005), and time-resolved connectivity, supporting 4D fMRI data (e.g. https://www.youtube.com/watch?v=yrjTGc5gkGo).

**Invasive electrodes:** The pipeline for invasive electrodes can identify electrodes' contacts in the co-registered CT, group them to reconstruct the depth electrodes, and "snap" them on the dural surface to reconstruct the actual post-implantation surface configuration, *Figure 2*, part (5). The electrode locations can be used to identify their associated gray-matter source and calculate connectivity. In more detail, the pipeline includes the following steps:

1. Identify stereotactic EEG (sEEG) electrodes' contacts from the post-implantation CT and group them to different electrodes using a semi-automatic algorithm (Peled et al., 2017). The coordinates of the electrodes' contacts are transformed from the CT acquisition space to the native space of the subject.
2. Import the raw sEEG recording data into MMVT in European Data Format (EDF), which can be exported from all commercial and research EEG and invasive electrodes acquisition systems. The specific time-window that represents the desired event, e.g., an epileptic seizure or interictal spike, can be selected on the native acquisition system.
3. Transfer the coordinates of the electrode to a template brain and/or between subjects using the FreeSurfer combined volumetric and surface registration (Postelnicu et al., 2009; Zöllei et al., 2010).
4. Transform electrodes to positions on the dural surface using simulated annealing "snapping" algorithm, which minimizes an objective energy function (Dykstra et al., 2012). This step is crucial for electrocorticography (ECoG) using electrode strips and grids. We have addressed this problem in a standalone tool (LaPlante et al., 2017).

5. Estimate the probability that a particular brain region contributes to the source of the signal at each electrode by using the Electrode Labeling Algorithm (ELA) (Peled et al., 2017). This algorithm operates under the assumption that the probability is estimated based on the Euclidean distance between the electrode and the brain labels (given a specific cortical atlas).
6. Calculate the frequency-domain and time-frequency-domain electrode connectivity based on estimates of the CSD and PSD.
7. Real-time streaming of the electrodes' waveforms into the MMVT GUI via UDP.

*2.3 MMVT interaction*

MMVT allows users to interact with the software's methods and tools in three primary ways: through the graphical user interface (GUI), see *Figure 3*, calling objects from a script using a "proxy object", and by Jupyter Notebook widgets. An example of each of the interaction options is depicted as follows.

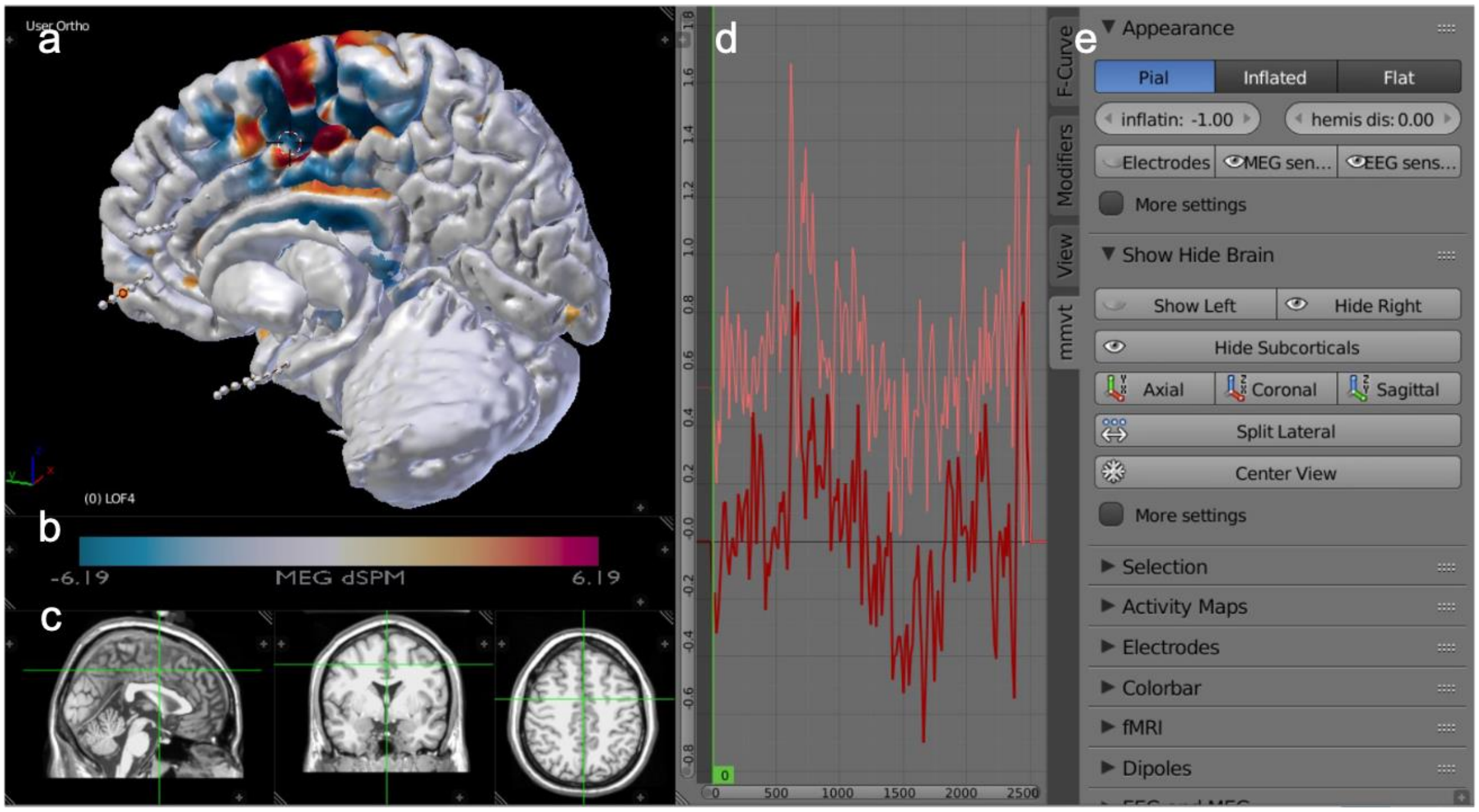

**Figure 3. Main MMVT GUI. (a)** A 3D brain view, **(b)** Color-bar that are updated automatically according to the activity being plotted, **(c)** Slices viewer for MRI (T1, T2, and FLAIR) and CT, **(d)** Time-domain and frequency-domain graphs, and **(e)** MMVT panels and buttons.

**Employ a "proxy object" to call MMVT objects from a script:** A second method to interact with the tool is to import the MMVT library with user-written Python scripts, as shown in Supplementary Figure S1. In this example, the user calls the MEG preprocessing pipeline to calculate the inverse operator and estimate the sources, based on the MNE-Python example dataset. Then, MMVT main GUI is opened to plot the calculated activity at a given time point and threshold.

The software supports use with Jupyter Notebook, where not only methods from the library can be called, but users can also create and embed figures from within the notebook or create an interactive viewer. Supplementary Figure S2 is an example of a source estimate light viewer. In this example, where we present the main part of the code, the user can choose a source estimate file, the time and perspective, and visualize the activity. There is also an option to show a movie of the activity and to render a figure. When the user changes the time, threshold, or perspective, there is a call to MMVT to produce a figure which is automatically loaded into the web-viewer. *MMVT can run in the background on the native computer or on a remote server. The procedure is very fast and can be completed in real-time.*

**Jupyter notebook widgets and macros:** These are Python scripts with a specific signature that can be run inside MMVT from the "Scripts" panel. This option is especially useful when running the same commands in the MMVT GUI for several subjects. An example of such usage is the reports generator, which generates a report within MMVT using the same series of actions for each subject to create the necessary figures. One can create a script that generates those figures, creates pdf files, and opens them. This is often useful for clinical studies, where physicians want an identical report for each of their patients. Another option is to send an MMVT file with a script that generates the desired results while working with the GUI to increase reproducibility.

## Results

In this section, we discuss five use cases to illustrates the added value MMVT gives to multimodal data exploration. The code for generating and plotting use cases #2 and #4 can be found in the GitHub repository of MMVT (Peled and Felsenstein, 2017)

*Use case #1: Estimate the Sources of TMS-Evoked Potentials Recorded with EEG*

We used MMVT to visualize neural activity in the study of consciousness. We combined Transcranial Magnetic Stimulation (TMS) and EEG acquisition in healthy, awake controls to investigate the spread of a TMS-evoked potential in the source space. By estimating the distribution of the neural currents as a function of time following single-pulse TMS, we can begin to understand the neural mechanisms of consciousness by studying both healthy controls and different psychiatric populations. We recorded EEG with 64 electrodes and translated their locations, as well as the location of the TMS coil, to a common MRI-coordinate system. We can then view the distributed current estimate, as well as the sensor-level signals as a function of time, to visualize the evolution of evoked brain activity following a TMS pulse, see Figure 4. The user can also pick a single electrode to see the evoked activity from a particular channel in comparison to the source localization activity of the corresponding brain region. In the future, we plan to use MMVT in real-time during TMS-EEG clinical sessions. Being able to estimate the sources of the evoked potential from the TMS pulses simultaneously during the sessions would allow us to ensure that the TMS pulses are delivered to the correct target and begin to understand the data already during the acquisition.

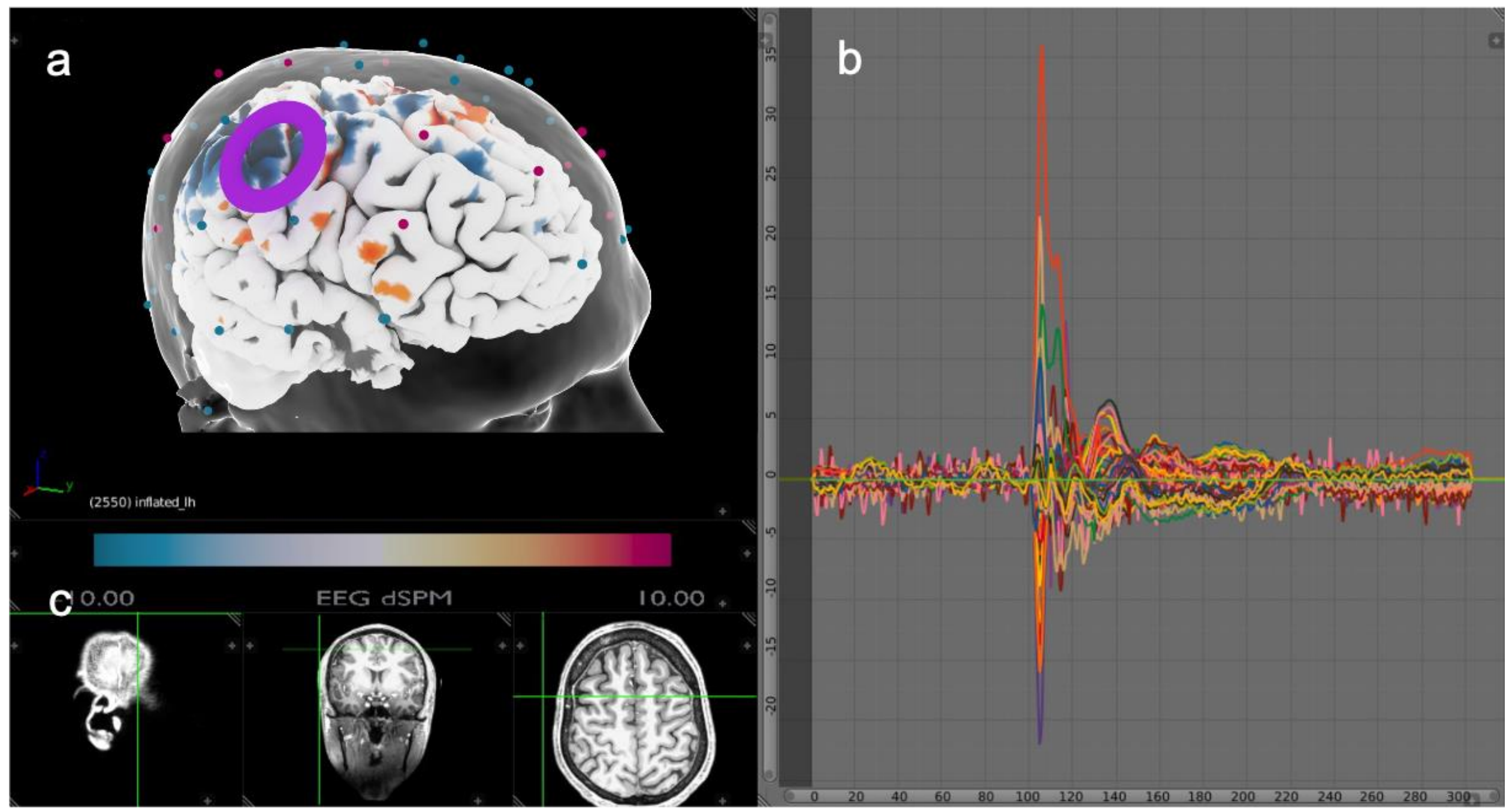

**Figure 4. EEG-TMS study (a)** Source localization (via dSPM) of neural activity resulting from the TMS-pulses, where the red EEG sensors were showing the greatest effect. The TMS coil is plotted as the blue torus. **(b)** Butterfly plot of the evoked responses of the EEG sensors. **(c)** The location of the TMS coil on the subject's MRI slices

*Use Case #2: Compare Resting-State Connectivity in fMRI and MEG*

To compare resting-state connectivity between fMRI and MEG, we calculated the connectivity matrix of 28 healthy subjects for each modality. The cortical surfaces were parcellated using the Lausanne125 atlas (Daducci et al., 2012), giving 219 x 219 x #windows matrix (we removed the corpus-callosum). Then, the sliding windows approach was utilized: for fMRI, the length is 34 TR, and a shift of 3 TR, where the TR was 3s. For MEG, the length of the window was 1 second and a shift of 0.5 seconds). For connectivity measure, we used the correlation coefficient for fMRI (Van Dijk et al., 2009), and the unbiased estimator of squared PLI (Vinck et al., 2011) for MEG. These results were averaged across subjects and windows. We transformed the connectivity matrix to a connectivity degree by using a cutoff at the 90$^{th}$ percentile and by counting the connections above this cutoff. Using the connectivity degree matrix, we found the main hub in each modality (*Figure 5*). The cortical labels with the highest numbers of connections included: right superiorfrontal-7 for fMRI (right dmPFC, 21.33, -0.12, 69 in MNI305) and left precuneus-2 for MEG (-10.92, -43.36,48.62 in MNI305).

Each hub was plotted (fMRI in red and MEG in blue) with its 15 strongest connections. Moreover, we plotted the connectivity degree matrix on a flat map. To visualize the results, we merged the two connectivity matrices into one and imported the combined matrix and the fMRI and MEG matrices into MMVT using the connectivity panel.

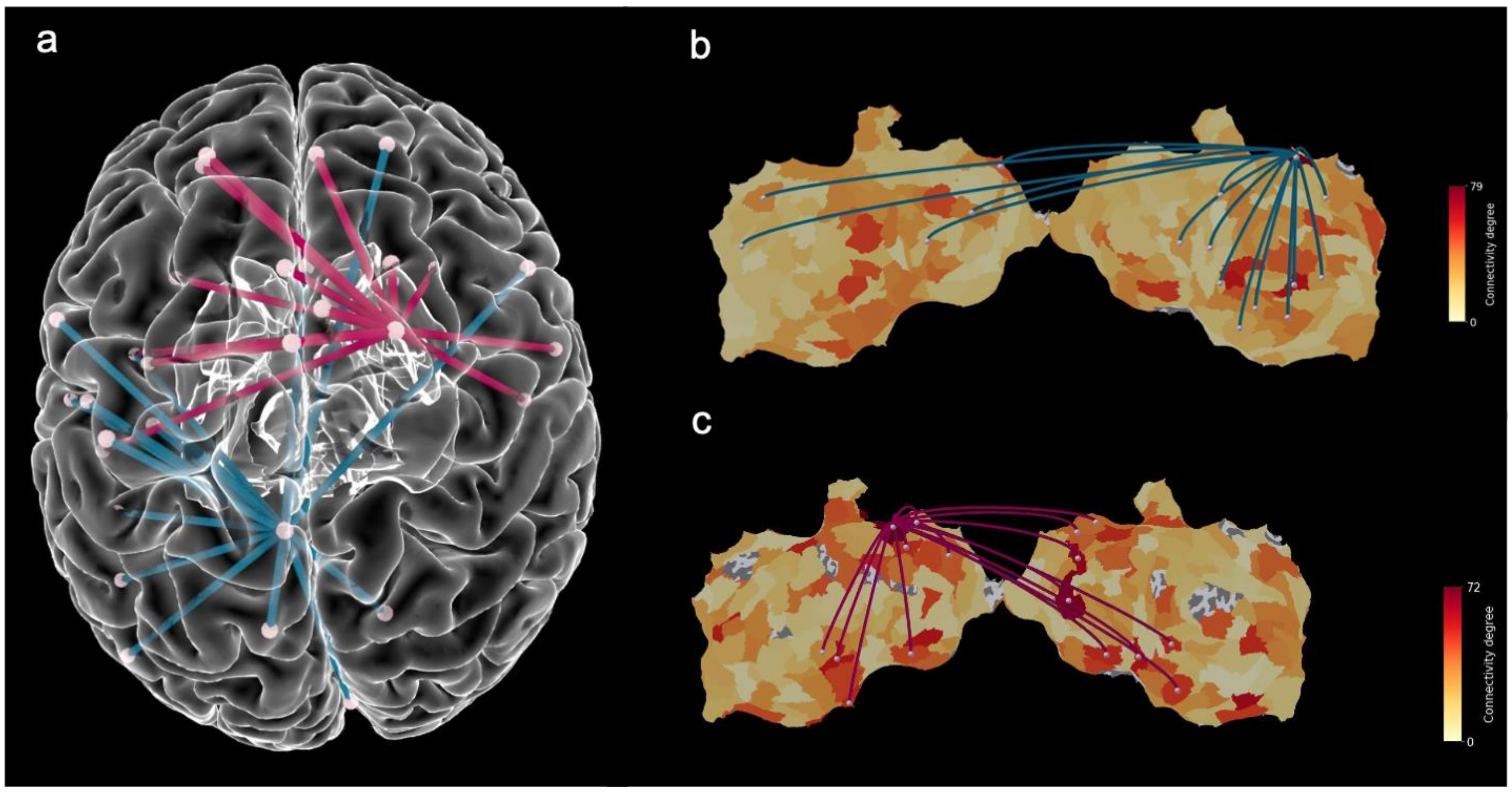

**Figure 5. Comparing the main connectivity hubs between MEG and fMRI. (a)** Both main hubs in fMRI (right dmPFC, in red) and MEG (left precuneus in blue) are marked in green. **(b)** MEG and **(c)** fMRI main hubs connectivity on the flat brain, where the colors represent the degree of connectivity, parcellated using the Lausanne125 cortical atlas.

*Use Case #3: Identification of Invasive Electrode Locations using MRI and CT*

We co-registered a patient's CT and MRI, to visualize sEEG electrodes in the patient's MRI space and show the two task-evoked responses for the MSIT task (Bush and Shin, 2006) from one of the electrodes. First, we co-registered the patient's CT to their MRI. To view the CT (as blue elements) on top of the MRI in slices viewer (*Figure 8b*), we found all the voxels in the CT above the 99.5 percentile (mostly the electrodes and the skull) and used the CT to MRI coordinate transformation to insert those values into the MRI slices. After the co-registration, we used a semi-automatic algorithm to identify the electrode contacts and to group them to reconstruct the depth electrodes shown in a unique color (Peled et al., 2017). We also created the FreeSurefer reconstruction from the MRI for visualization. By selecting an electrode contact the slice viewer was automatically updated and the task-evoked responses are shown in the graph panel (*Figure 6c*), where dark and light red distinguished the two conditions of the task. Using the slicer panel, we created a superior axial cut. The axial slice including both MRI and elements from the CT were plotted on the 3D slice. The electrodes are visualized as small blue circles both in the slice viewer and around the red electrodes in the 3D slice.

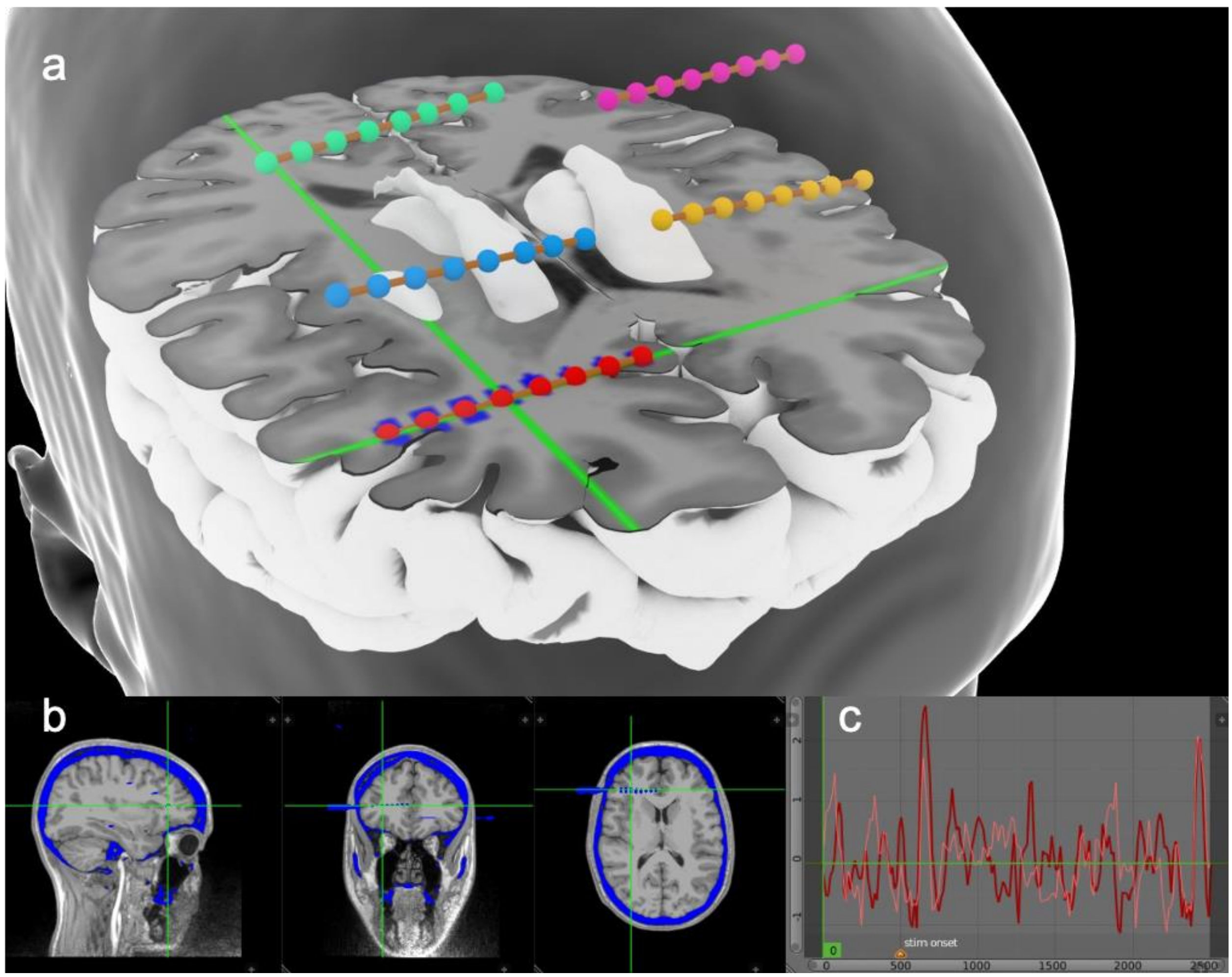

**Figure 6. Depth electrodes co-registration. (a)** Combined 3D and transversal slice and **(b)** traditional orthogonal slices of a patient's brain in native space with depth electrodes that were placed based on the co-registration of pre-surgical MRI and post-surgical CT scan. Blue filled regions are voxels with high intensity in the CT. **(c)** The graph shows the evoked response from the selected electrode (green crosshair in the 3D viewer). The blue filled spaces in **(b)** are voxels with high intensity in the CT.

*Use Case #4: fMRI Activity and MEG Gamma Power*

We examined the correlation between fMRI BOLD signal and MEG gamma power across time. Previous work has shown significant correlations between both low and high gamma power and fMRI BOLD in a visual task-based analysis (De Pasquale et al., 2010; Muthukumaraswamy et al., 2009). We analyzed and compared results from the Multi-Source Interference Task (MSIT) task (Bush and Shin, 2006), Figure 7*a*, a Stroop-like task (Williams et al., 1996) that can be used to identify the attention network in healthy participants, and test its integrity in people with neuropsychiatric disorders. The MEG power spectrum was calculated for both MSIT's conditions (congruent and incongruent), and a Welch's t-test was used to calculate the p-values for each vertex. The -log10 of the p-values is plotted in the right panel of Figure 7*b* over frequencies. The peak of 78Hz was selected (vertical green line), and the values were plotted using the hot color-bar. For fMRI, two contrasts of interest were specified in a simple subtraction design. As shown, the highest correlation is in the left superior frontal gyrus, around (15.1, 60.2, 15,24) in MNI305 coordinates (Figure 7c*)*. This example demonstrates how such rich datasets can be explored in detail rather than just selecting a peak in the gamma range. For example, different frequencies can be selected from the graph panel to cause a replot of the gamma p-values on the cortical surfaces

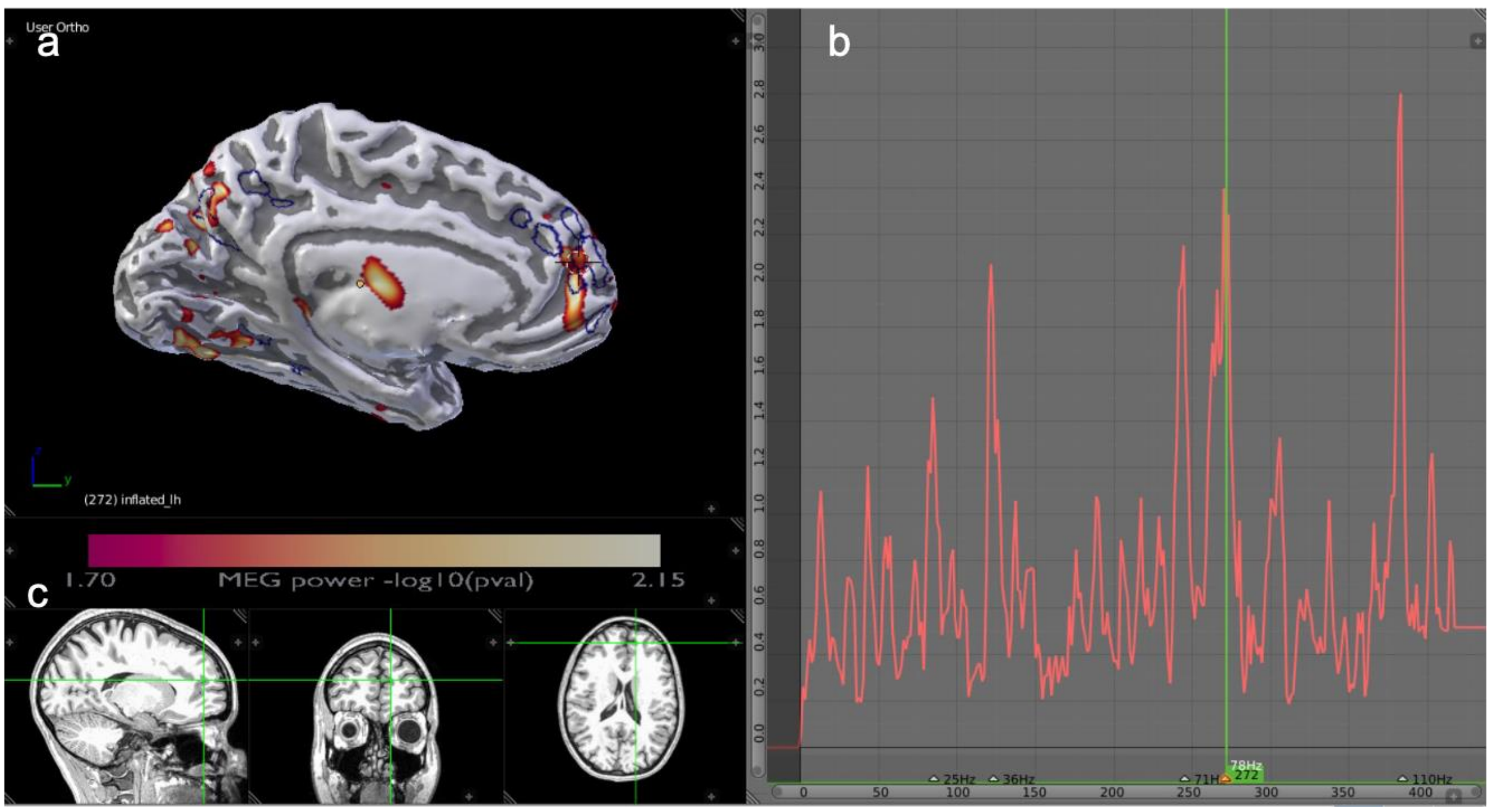

**Figure 7. fMRI hits and MEG high-gamma power. (a)** Correlation is in the left superior frontal between the MEG high-gamma Welch's t-test results in red-yellow colors, and the fMRI contrast map in blue contours. **(b)** Welch's t-test MEG results over frequencies, where the 78 Hz is selected. **(c)** The location of the intersection in the slices viewer

*Case study #5 PET DBS study*

In this study, inter-subject PET imaging analyses were visualized with DBS electrode locations. The aim is to quantitatively assess and visualize longitudinal changes in glucose metabolism affected by deep brain stimulation. Inter-subject analyses were performed given the quantitative limitations of the available static-acquisition intra-subject data (Yoder, 2013). There are several advantages to performing the co-registration of PET and MR modalities in MMVT. MMVT can choose virtually any MRI or PET brain templates, toggle between atlases to determine which template best fits the data provided, provide an advanced visual inspection of co-registration in cortical and subcortical regions, and visualize intra-session changes given dynamically acquired data. These quality control processes help determine the specificity of activation limited by precision with respect to anatomical landmarks, especially where subcortical regional activation is involved. To demonstrate that MMVT can enhance specificity, we present preliminary data from a DBS study targeting the ventral capsule/ventral striatum. As a case study, we show PET scans from one patient with intractable OCD three months post-implantation of a Ventral Internal Capsule/Ventral Striatum deep brain stimulation device. To visualize the PET results on the template brain, we transformed the location of one of the patient's DBS electrodes from the patient's native MRI space to the template space, as can be seen in Figure 8.

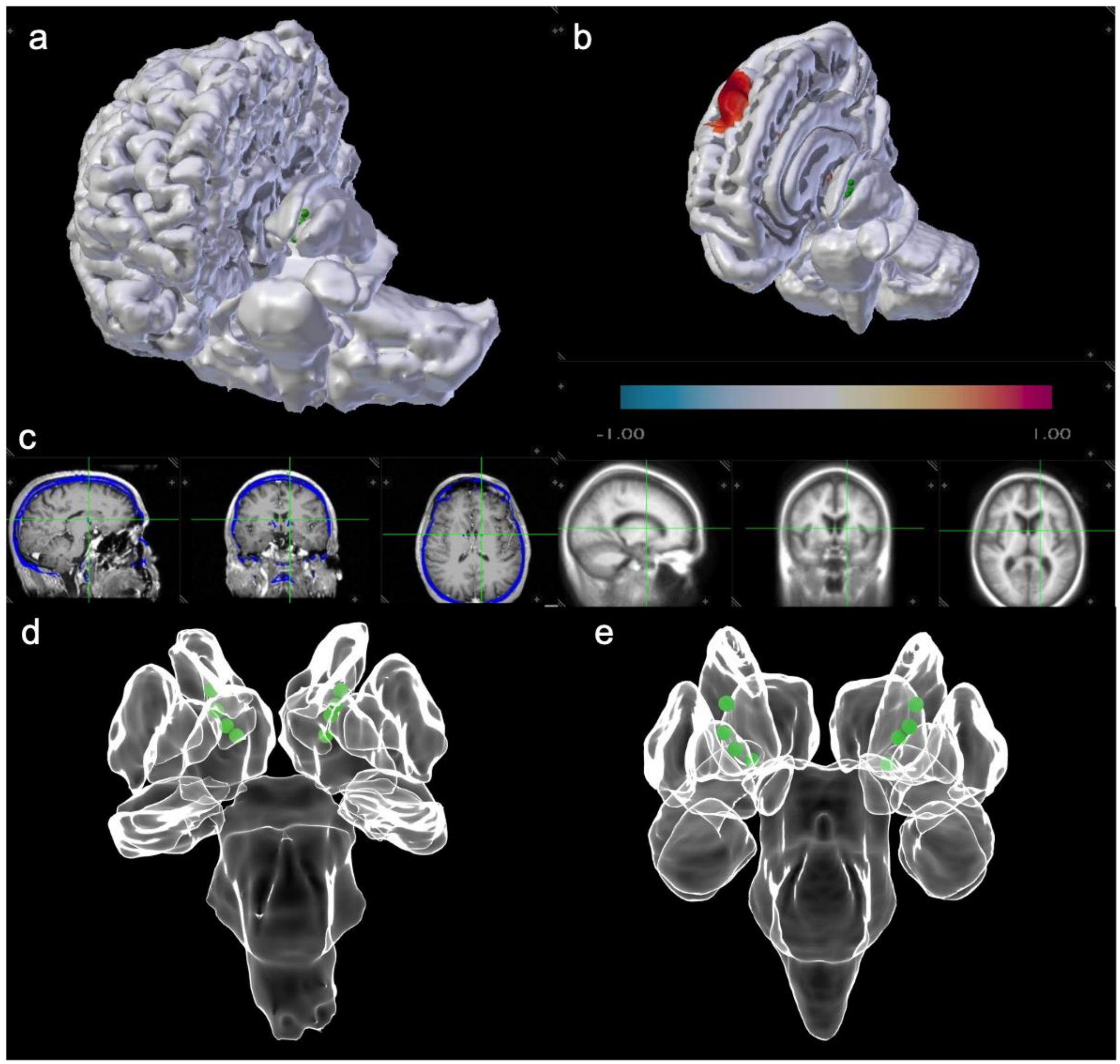

**Figure 8. Electrodes morphing and PET activation.** The electrodes from the right hemisphere are shown (as green spheres) in **(a)** the patient's native MRI space and the **(b)** template brain, where the PET group analysis results are shown in red. **(c)** The selected electrode's location in the slices viewer. Blue voxels in the slices are elements with high CT intensity, like the skull and the DBS electrodes. To visualize better the electrodes' location, we rendered a transparent sub-cortical view of the DBS electrodes **(d)** in the patient's native MRI space, and **(e)** in the template brain

## Discussion

The MMVT tool consists of two main parts: an easy to use data preprocessing tool with modules and analysis pipelines implemented in Python, and a graphic user-interface with the ability to bring multiple modalities into the same user-friendly space for visualization. Researchers can increase their analysis efficiency by harnessing the richness of data exploration using the graphical interface without losing the strengths of open-source design. The software is released as open-source software under GNU General Public License v3.0, allowing researchers to seamlessly integrate desired aspects of MMVT into their pipelines for exploration or data visualization.

Researchers and clinicians can greatly benefit from the multimodal visualization tool, as demonstrated by the five case studies. Data acquired on multiple modalities using different commercial imaging or neurophysiology packages are only useful when reviewed together on the same screen. Interpretation of intracranial imaging and correlating it to possible surgical or neurostimulation targets can significantly change the yield of the studies, and in turn treatment outcome (Bartolomei et al., 2018). Communication between team members is facilitated by clearly showing the overlapped multimodal images. This could eliminate confusion related to localization of brain regions of interest. Beyond visualization, the streamlined and robust pre-processing tools provide clinicians with reformatted, ready to use data in native space. This data can be in turned analyzed using any additional desired techniques without the hassle of adopting a step-by-step pre-processing.

In clinical neuromodulation such as DBS, programming often requires a search through a large, non-linear space (Widge et al., 2019). The current best practice relies heavily on patient reports, which can be difficult to interpret. There is a long delay between stimulation settings and symptom change, but the necessary brain changes may be more quickly seen in electrophysiology (Holtzheimer and Mayberg, 2011; Widge et al., 2018). MMVT visualization of DBS and LFPs in real-time gives clinicians vastly greater information on the effects of their changes in the stimulation parameters on the effected brain circuitry. MMVT can make the parameter search far more efficient by, for example, uploading the results obtained for a range of values for a specific parameter and exploring the spatial organization obtained by each value.

In basic research, studies employing multiple imagining modalities would benefit from a streamlined pipeline that allows them to quickly and intuitively integrate their findings, and present them simultaneously (Calhoun and Sui, 2016). This is of special interest and importance for modalities that capture different spatial and temporal resolutions, such as M/EEG and fMRI. Overcoming such an issue, MMVT enables picking an activity cluster from the fMRI analysis to create a functional ROI within which the averaged M/EEG activity can be explored. By seemingly integrating data across modalities into a common space (normalised or native), MMVT provides the user with flexibility to explore and visualize complex relationships in space, time, and frequency.

The real-time streaming capabilities of MMVT will enable brain stimulation researchers to visualize stimulation-evoked potentials in real-time. In TMS-EEG, small changes in position and orientation of the coil in relation to the geometry of the scalp, as small as 1 mm, produce qualitatively different magnitude and characteristics of TMS-evoked potentials (Casarotto et al., 2010; Rosanova et al., 2009). MMVT can visualize TMS-evoked EEG potentials in real-time to precisely navigate to the optimal stimulation position and orientation, guided by individual brain morphology. Since MMVT is open-source, researchers have the flexibility to design innovative studies and to implement custom functionalities. For example, researchers could integrate computations of estimated TMS electric fields that could be visualized and compared to all other modalities based on the anatomical surfaces of individuals or group templates simultaneously.

In the coming years, we expect to see an increasing number of studies with combinations of neuroimaging techniques, as well as addition of brain stimulation techniques. This is due to recent advancements in performing TMS simultaneously with fMRI and EEG (George et al., 2019) as well as MR compatible electrodes and DBS probes (Guerin et al., 2018; Nimbalkar et al., 2019). These exciting new developments open new opportunities for both basic and clinical neuroscience. As this field continues to evolve, the intuitive and extensive MMVT platform could help uncover the relations between neural computations captured at different scales, as well as brain stimulation and brain mapping. The exploration capabilities of MMVT can be very powerful in investigating these new types of multi-dimensional and multi-modal data.

**Acknowledgments**

This research was funded by the Defense Advanced Research Projects Agency (DARPA) under Cooperative Agreement Number W911NF-14-2-0045 issued by ARO contracting office in support of DARPA's SUBNETS Program. The views, opinions, and/or findings expressed are those of the author and should not be interpreted as representing the official views or policies of the Department of Defense or the US Government.

This research was also sponsored by NCRR (S10RR014978) and NIH (S10RR031599, R01NS069696, 5R01NS060918, U01MH093765, 1R01NS104585, 5R01EB022889).

O. Felsenstein gratefully acknowledges the support and funding by the President's scholarship awarded by Bar Ilan University and the Israeli Center of Research Excellence in Cognition (I-CORE Program 51/11).

N. Peled and S. Stufflebeam are co-founders in FIND Surgical Sciences Inc.